\documentclass[a4paper,fleqn,usenatbib]{mnras}

\usepackage[T1]{fontenc}
\usepackage{ae,aecompl}

\usepackage{graphicx}	

\title[Radio properties of H$_2$O megamaser Seyfert2]{A systematic observational study of radio properties of H$_2$O megamaser Seyfert-2 galaxies}

\author[Z.W.~Liu et al.]{
Z.W.~Liu,$^{1}$
J.S.~Zhang,$^{1}$\thanks{E-mail: jszhang@gzhu.edu.cn}
C.~Henkel,$^{2,3}$
J.~Liu,$^{4}$
P.~M{\"u}ller,$^{2}$
J.Z.~Wang$^{5}$
\newauthor
Q.~Guo,$^{6}$
J.~Wang$^{1}$
and J.~Li$^{5}$
\\
\\
$^{1}$Center For Astrophysics, GuangZhou University, GuangZhou 510006, China\\
$^{2}$Max-Planck-Institut f{\"u}r Radioastronomie, Auf dem H{\"u}gel 69, D-53121 Bonn, Germany\\
$^{3}$Astronomy Department, King Abdulaziz University, P.O. Box 80203, Jeddah 21589, Saudi Arabia\\
$^{4}$Xinjiang Astronomical Observatory, CAS, 150, Science 1-Street, Urumqi, Xinjiang 830011, China\\
$^{5}$Shanghai Astronomical Observatory, CAS, 80, Nandan Road, Shanghai 200030, China\\
$^{6}$Hunan Institute of Humanities, Science and Technology, Loudi 417000, China
}

\date{Accepted XXX. Received YYY; in original form ZZZ}

\pubyear{2016}

\begin{document}
\label{firstpage}
\pagerange{\pageref{firstpage}--\pageref{lastpage}}
\maketitle

\begin{abstract}
A systematic study is performed on radio properties of H$_2$O megamaser host Seyfert 2 galaxies,
 through multi-band radio continuum observations (at 11\,cm, 6.0\,cm, 3.6\,cm, 2.0\,cm and 1.3\,cm) with the Effelsberg 100-m radio telescope within a total time duration of four days. 
For comparison, a control Seyfert 2 galaxy sample without detected maser emission was also observed. 
Spectral indices were determined for those sources for which measurements exist at two adjacent bands assuming a power-law dependence S$_\nu \propto \nu^{-\alpha}$, where  S is the flux density and $\nu$ is the frequency. Comparisons of the radio continuum properties between megamaser and non-masing Seyfert 2s show no difference in spectral indices. However, a difference in radio luminosity is statistically significant, i.e. the maser galaxies tend to have higher radio luminosities by a factor of 2 to 3 than the non-masing ones, commonly reaching values above a critical threshold of 10$^{29}$\,erg\,s$^{-1}$\,Hz$^{-1}$. This result confirms an earlier conclusion by Zhang et al. (2012), but is based on superior data with respect to the time interval within which the data were obtained, with respect to the observational facility (only one telescope used), the number of frequency bands.

\end{abstract}

\begin{keywords}
Masers -- galaxies: active -- galaxies: nuclei -- radio lines: galaxies --
radio continuum: galaxies
\end{keywords}



\section{Introduction}

Great efforts have been made to study H$_2$O masers in extragalactic systems at $\lambda$ $\sim$ 1.35\,cm (22.23508 \,GHz) in the J$_{\rm KaKc}$ = 6$_{16}$-5$_{23}$ transition, since the first detection of such an extragalactic maser towards M33 \citep{1977A&A....54..969C}. To date, the maser line has been detected in over 160 galaxies (see the Megamaser Cosmology Project (MCP), https://safe.nrao.edu/wiki/bin/view/Main/Megamaser CosmologyProjectwebpages). These masers could be categorized by two classes to be associated with: (1) star formation regions (2) active galactic nuclei (AGN)
(e.g. \citealt{2005Ap&SS.295..107H}; \citealt{2006A&A...450..933Z}).
Among AGN-related masers, more than 30 sources have been identified as "disk-maser" candidates (\citealt{2006ApJ...652..136K,2010ApJ...708.1528Z}; \citealt{Greene:2016wv}; the MCP webpage). The maser features in "disk-maser" systems trace a thin, edge-on Keplerian disk on sub-parsec scales around the central supermassive black hole (SMBH), which provides an excellent  tool for accurate determinations of the black hole mass and the Hubble constant.

Observations and studies show that extragalactic H$_2$O masers with an isotropic luminosity greater than 10\,L$_\odot$, which are termed "megamasers", are mostly found in galaxies that are categorized as Seyfert 2 or LINER (Low Ionization Nuclear Emission-Line Region) galaxies. These are heavily obscured with gas column densities N$_H$ $>$ 10$^{23}$cm$^{-2}$ \citep{1997ApJS..110..321B,2006A&A...450..933Z,2008ApJ...686L..13G}. A small number of sources which have been studied with high spatial resolution strongly indicate that all megamasers are AGN related. Population inversion for the H$_2$O 6$_{16}$-5$_{23}$ line can be explained by collisional pumping with the AGN being considered to be the ultimate energy source that feeds the maser emission \citep{Lo:2005iq}. Potentially required "seed" photons for the maser medium aligned with the nucleus or associated with a nuclear jet may be provided by radio emission from the nucleus or jet, which would then be amplified by the maser medium, leading to strong detectable H$_2$O profiles. On the other side, the isotropic luminosity of the nuclear radio continuum is believed to be an indicator of AGN power \citep{1990ApJS...72..551G,2009ApJ...698..623D}. Thus we expect some kind of correlation between the isotropic luminosities of megamasers and the nuclear radio emission. 

This was firstly investigated by \citet{2012A&A...538A.152Z}. Based on collected data at 20\,cm and 6.0\,cm, they have proposed that maser host galaxies have higher nuclear radio continuum luminosities, exceeding those of a comparison sample by factors of order 5. Therefore nuclear radio luminosity was suggested to be a suitable indicator to guide future AGN maser searches. However, the uncertainties of that analysis are still quite high. 
For both maser galaxies and non-masing galaxies, measured radio data commonly come from different telescopes.
Even if data could be taken from the same telescope, measurements were normally performed at different epochs.
In addition, there are just a few data at other radio bands for maser host galaxies, e.g., 3.6\,cm, 2.0\,cm. So a presentation of a more complete dataset is urgently needed. While an interferometric study would be a better choice, systematic studies of the lower resolution radio continuum with single-dish telescopes are still worthwhile, because they are not affected by missing flux and can guide future megamaser surveys. Thus we conducted systematic Effelsberg multi-band observations toward the H$_2$O megamaser host galaxies and a control galaxy sample devoid of detected maser emission.

\section{The sample}

As mentioned above, most maser host galaxies are Seyfert 2s or LINERs. Thus Seyfert 2 galaxies from the megamaser sample and a control sample are chosen to be our targets.
Among 85 published H$_2$O maser galaxies, there are 49 Seyfert 2 sources.
For the control Seyfert galaxy sample, we consider a complete sample consisting of 89 relatively nearby Seyfert galaxies compiled by \citet{2009ApJ...698..623D}, which was drawn from the revised Shapley-Ames (RSA) catalog that selects galaxies having B$_T$ < 13.31\,mag \citep{1995ApJ...454...95M, 1997ApJS..112..315H}. 
This control Seyfert sample is unique, in that it has been searched for H$_2$O maser emission. 54 of the 89 galaxies are Sy2s, which provide a good comparison sample (see details in \citealt{2012A&A...538A.152Z}). 
Among both the megamaser (49 sources) and the nonmaser (54 sources) Seyfert 2 sample, 35 masers and 25 non-masing sources with Declination >\,-20 degrees were observed with the Effelsberg telescope. All 35 maser sources are AGN-related and 15 of them are possibly disk-masers \citep{2010ApJ...708.1528Z,2011ApJ...727...20K,2016AAS...22724359P}.

\section{Observation and data reduction}

The multi-band radio continuum (11\,cm, 6.0\,cm, 3.6\,cm, 2.0\,cm and 1.3\,cm) observations were performed on 18 to 21 January, 2014, with the Effelsberg 100-m radio telescope of the Max-Planck-Institut f{\"u}r Radioastronomie (MPIfR), using secondary focus heterodyne receivers. 
All of the flux measurements were done in cross-scan mode, where the antenna beam pattern was driven repeatedly in azimuth and elevation over the source position, switching between our targets and calibrators. Our targets were observed one by one and each source was observed in all bands within one hour, with a typical total interval time of $\sim$ 40 minutes.

\begin{figure}
\centering
\includegraphics[width=\columnwidth]{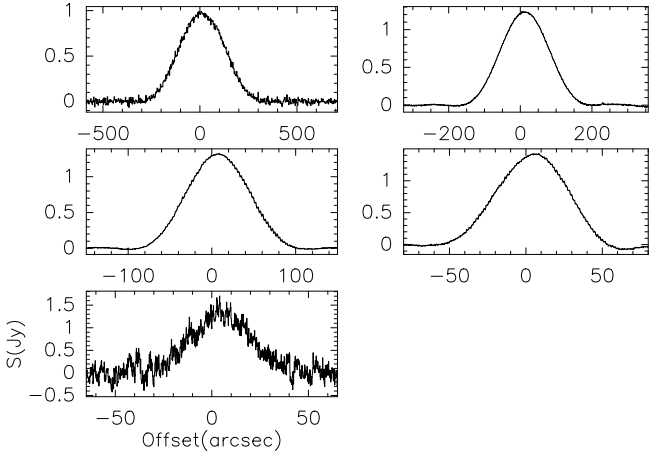}
\caption{Continuum cross scans of NGC1052 at 11\,cm and 6.0\,cm (top), 3.6\,cm and 2.0\,cm (intermediate) and 1.3\,cm (bottom), respectively. For the other sources, see the 
            Appendix.}
\label{figure0}
\end{figure}

The calibration for each wavelength was obtained with the following procedure. First, baseline subtraction and Gaussian profile fitting were performed to each individual sub-scan. The amplitude of the profile provides an estimate of the source's flux density expressed in units of antenna temperature (K), divided by the signal from the noise diode. Then the Gaussian amplitude, offset and half power beam width (HPBW) of the sub-scans were independently averaged in each driving direction. The typical pointing error for the Effelsberg radio telescope is $\sim$\,2 arcseconds,
 which will result in a flux density underestimation of $\sim$2\%. However, this can be corrected by pointing correction which assumes a two dimensional Gaussian intensity profile.
Subsequently, the pointing-error corrected amplitudes from both scanning directions were averaged together,  providing a single flux density measurement per scan. After this, an opacity correction was made for each scan deduced from the obtained system temperature (T$_{sys}$). Next the systematic elevation-dependent gain variations were corrected by a polynomial function derived from non-variable calibrator sources (e.g., 3C48, 3C286, 3C295 and 3C138). Then the time-dependent gain fluctuations, which are mainly caused by changing weather conditions, were corrected by a gain-time transfer function obtained from measurements of several secondary calibrators. Finally the measured antenna temperatures (in K) of each source were converted to flux densities (in Jy), by using primary calibrator measurements (e.g. \citealt{1977A&A....61...99B}; \citealt{1994A&A...284..331O}).
The final flux density uncertainty is composed of the statistical errors from the Gaussian fit, the weighted average over the sub-scans, the gain and time-dependent corrections, and a contribution from the residual scatter seen in the primary and secondary calibrator measurements, which characterizes uncorrected residual effects. The relative error seldom reaches up to 10\% for target sources and 2\% for calibrators.
Figure \ref{figure0} shows one example characterizing our sample of observed sources.The profiles of cross-scans of other sources are shown in the Appendix.

The above described cross-scan data calibration technique is well established and allows for high precision flux density determination(e.g. \citealt{2003A&A...401..161K}; \citealt{2008A&A...490.1019F}).

\section{Analysis and discussion}
\subsection{Comparison of radio luminosities of H$_2$O maser and non-masing galaxies}

\begin{table*}
\caption{The comparison of radio properties of megamaser and non-masing Seyfert 2s}
\label{table3}
\centering
\resizebox{1\textwidth}{!}{%
\begin{tabular}{ccccccccccc}
\hline 
Samples & Subsamples &  $\log$ \emph L$_{11}$ & $\log$ \emph L$_{6.0}$ & $\log$ \emph L$_{3.6}$ & $\log$ \emph L$_{2.0}$ & $\log$ \emph L$_{1.3}$ & $\alpha_{6.0}^{11}$     &  $\alpha_{3.6}^{6}$ 	& $\alpha_{2.0}^{3.6}$ 	& $\alpha_{1.3}^{2.0}$\\ 
 &  & \multicolumn{5}{c}{(erg$\cdot$s$^{-1}$Hz$^{-1}$)} &	&	&	&	\\ \hline
Maser 	   & Total 		      &  29.49 $\pm$ 0.02 	& 29.18 $\pm$ 0.01 	& 29.02 $\pm$ 0.02	& 29.35 $\pm$ 0.04 	& 30.09 $\pm$ 0.03	&  1.02 $\pm$ 0.14 &  0.96 $\pm$ 0.11 & -0.12 $\pm$ 0.26 & -0.27 $\pm$ 0.31	\\
 		   & D\textless70Mpc  &  29.34 $\pm$ 0.01   & 29.05 $\pm$ 0.01  & 28.92 $\pm$ 0.02  & 29.27 $\pm$ 0.03  & 30.09 $\pm$ 0.03		& 0.95 $\pm$ 0.13 & 0.95 $\pm$ 0.14 & 0.24 $\pm$ 0.19 & -0.27 $\pm$ 0.31 \\  
non-masing & Total            &  28.71 $\pm$ 0.02   & 28.35 $\pm$ 0.02  & 28.35 $\pm$ 0.03  & 28.41 $\pm$ 0.03  & - &  1.01 $\pm$ 0.16 &  1.01 $\pm$ 0.17	& -0.02 $\pm$ 0.25	&  - \\ 
 \hline 
 t-Test Prob. 	& Total             	& <\,0.001 & <\,0.001   & 0.006      & 0.013   & -  & 0.787 &  0.788	& - & - \\    
   			    & D\textless\,70Mpc 	& 0.007    & 0.003      & 0.027      & 0.068   & -  & - & - & - &- \\	  \hline
 Logrank-Prob.  & Total            		& <\,0.001 & <\,0.001   & 0.009      & 0.170   & -  & 0.178 &  0.457	& -	& -	\\ 
   			    & D\textless\,70Mpc 	& 0.007    & 0.002	    & 0.049      & 0.420   & -  & - & - & - &- \\    \hline
\end{tabular}
}
\end{table*}

Among the 35 H$_2$O megamaser Seyfert 2 galaxies, 24 sources ($\sim$\,68.6$\%$) are detected at 11\,cm, 25 sources ($\sim$\,71.4$\%$)at 6.0\,cm, 21 sources (60$\%$) at 3.6\,cm, 10 sources ($\sim$ 28.6$\%$) at 2.0\,cm and 2 ($\sim$ 5.7$\%$) at 1.3\,cm. Among the 25 non-masing Seyfert 2s, 16 sources (64.0$\%$) are detected at 11\,cm, 17 sources (68$\%$) at 6.0\,cm, 11 sources (44.0$\%$) at 3.6\,cm, and 4 sources (16.0$\%$) at 2.0\,cm while no sources at 1.3\,cm have been measured with signal-to-noise ratios larger than three (for details, please refer to Tables \ref{masergalaxies} and \ref{nonmaser}, upper limits are given for the flux density of those undetected sources). 

The corresponding luminosity of each individual source at each band, assuming here and elsewhere isotropic emission, are calculated by
\begin{equation}
	L_{\nu} = 4 \pi D^{2} S_{\nu}\,,
\end{equation}
where $D$ is the distance of a source and $S_{\nu}$ is the flux density. Assuming $H_0 = 70$\,km\,s$^{-1}$\,Mpc$^{-1}$, $\Omega_{\rm M} = 0.270$ and $\Omega_{\rm vac} = 0.730$ \citep[e.g.,][]{2003ApJS..148..175S}, the distance is calculated with Cosmology Calculator I provided by the NASA Extragalactic Database \citep{2006PASP..118.1711W}. The mean luminosities at each band on logarithmic scales (hereafter $\log L_{11cm}$, $\log L_{6.0cm}$, $\log L_{3.6cm}$ and $\log L_{2.0cm}$; these are given in units of erg$\cdot$s$^{-1}$Hz$^{-1}$) are listed in Table \ref{table3}. 

Below, we first assume that the galaxies of the maser and the control samples are, with respect to all their properties except maser and radio continuum luminosities, identical.
After comparing continuum luminosities of masing and non-masing targets in this way, we will take a deeper look into the properties of the two galaxy samples and will evaluate in how far potential differences will modify our previously obtained conclusions.

The mean radio continuum luminosities for maser Seyfert 2s that are detected are $\log L_{11cm}$ = 29.49\,$\pm$\,0.02, $\log L_{6.0cm}$ = 29.18\,$\pm$\,0.01, $\log L_{3.6cm}$ = 29.02\,$\pm$\,0.02, $\log L_{2.0cm}$ = 29.35\,$\pm$\,0.04 and $\log L_{1.3cm}$ = 30.09\,$\pm$\,0.03, and those of the non-masing Seyfert 2s are $\log L_{11cm}$ = 28.71\,$\pm$\,0.02, $\log L_{6.0cm}$ = 28.35\,$\pm$\,0.02, $\log L_{3.6cm}$ = 28.35\,$\pm$\,0.03 and $\log L_{2.0cm}$ = 28.41\,$\pm$\,0.03, respectively (throughout the paper, given errors are the standard deviations of the mean). By comparison, the H$_2$O maser sources tend to possess larger mean radio continuum luminosity, with a luminosity ratio of about 6.0\,$\pm$\,1.4, 6.8\,$\pm$\,1.6, 4.7\,$\pm$\,1.1 and 8.7\,$\pm$\,2.0 at 11\,cm, 6.0\,cm, 3.6\,cm and 2.0\,cm, respectively.
Furthermore, Welch's t-Tests were used here to check the difference of luminosity means between the entire maser and non-masing Seyfert2 samples at each band. The Welch's t-Test results (see details in Table \ref{table3}, row 4) show that the difference of the luminosity means is significant at 11\,cm, 6.0\,cm and 3.6\,cm, with a chance probability less than 0.05 (at 2.0\,cm, the non-masing sample of four sources is too small to provide a reliable statistical result).

\begin{figure}
\centering
\includegraphics[width=\columnwidth]{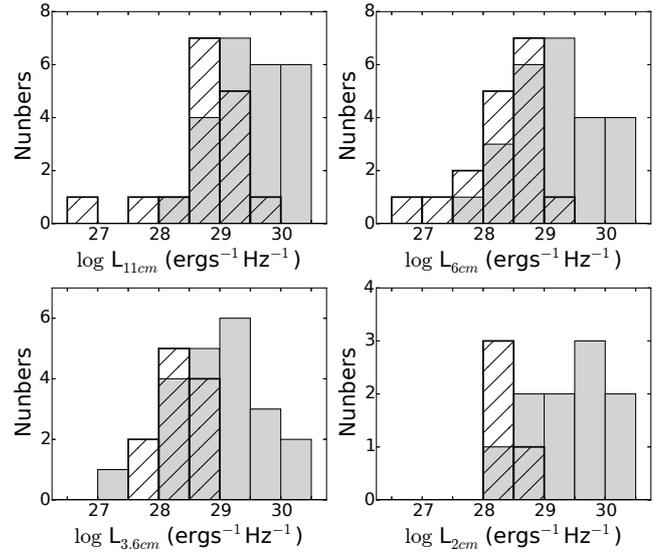}
\caption{Distributions of radio continuum luminosities (logarithmic scales, in units of erg$\cdot$s$^{-1}$Hz$^{-1}$) at 11\,cm (upper left panel), 6.0\,cm (upper right), 3.6\,cm (lower left) and 2.0\,cm (lower right), for Seyfert 2s (grey: maser sources; open histograms with diagonal lines: sources without detected masers).}
\label{4charc}
\end{figure}

The luminosity distributions of the four bands (11\,cm, 6.0\,cm, 3.6\,cm and 2.0\,cm) are plotted in Fig.\ref{4charc} for both the maser sample (in grey filled histograms) and the non-masing sample (open histograms with slashes), respectively, all with a logarithmic bin size of 0.5 dex, based on erg$\cdot$s$^{-1}$Hz$^{-1}$ units. A comparison between luminosity distributions of maser and non-masing samples at each of the four bands shows the same trend: maser host galaxies have larger radio luminosity, with an intersecting region present to some extent. Since upper flux density limits could be derived for undetected sources, survival analysis was performed here to investigate the difference in luminosity distributions. The logrank test results (listed in Table \ref{table3}) for the entire maser and non-masing sources show that the differences of the radio continuum luminosity distributions at 11\,cm, 6.0\,cm and 3.6\,cm are significant with a chance probability less than 0.05 in each case, while no reliable statistical result could be obtained from the 2.0\,cm data due to a too small sample.

\begin{figure}
\centering
\includegraphics[width=\columnwidth]{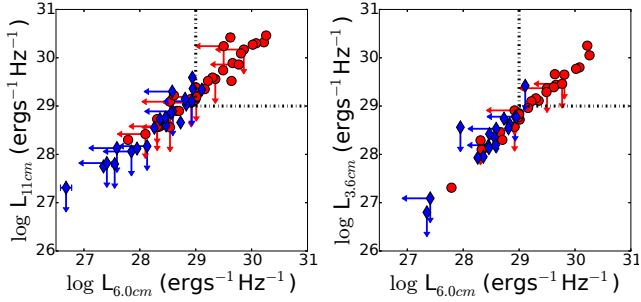}
\caption{Radio continuum luminosity (on logarithmic scales) at 6.0\,cm vs. 11\,cm (left panel) and 3.6\,cm (right panel). Red filled circles: maser sources. Blue filled diamonds: non-masing sources.}
\label{3lumvslum}
\end{figure}

The significant difference can also be visualized by Fig.\ref{3lumvslum}, which plots 6\,cm vs. 11\,cm (left panel) and 6\,cm vs. 3.6\,cm (right panel) luminosities for both maser(red circles) and non-masing (blue diamonds) Seyfert 2 sources, respectively. It demonstrates that most maser Seyfert 2 sources are located in the upper right region, at a higher luminosity, while the non-masing Seyfert 2 sources are located in the lower left, though there are a few overlaps. 
Together, the H$_2$O maser sources generally have higher radio luminosities than those of non-masing galaxies, almost an order of magnitude larger, meaning that H$_2$O megamasers are more likely to be found toward Seyfert 2 galaxies which possess larger radio luminosities. Dotted lines in Fig.\ref{3lumvslum} enclose those sources with radio luminosities $L_\nu\geq$\,10$^{29}$\,erg$\cdot$s$^{-1}$Hz$^{-1}$ simultaneously at the two bands (6.0\,cm and 11\,cm for the left panel; 6.0\,cm and 3.6\,cm for the right panel, respectively). All but one of them are megamaser sources in the limited region.
Nevertheless, we have to mention that the criterion $L_\nu\geq$\,10$^{29}$\,erg$\cdot$s$^{-1}$Hz$^{-1}$ is arbitrary, and the overlapping sections of the central regions of Fig.\ref{3lumvslum} suggest that there is no clearly defined boundary on radio continuum luminosities for distinguishing megamaser and non-masing Seyfert 2 galaxies. However, choosing specifically $L_\nu\geq$\,10$^{29}$\,erg$\cdot$s$^{-1}$Hz$^{-1}$ sources would drastically improve chances for detection.

Since the brightness of an object that is observed will decrease with its distance, sources at distances where the brightness falls below the observational threshold will be unobservable. The farther the celestial objects we observe, the higher the observational threshold, which results in a severe bias in statistical evaluations \citep{2012psa..book.....W}.
The mean values of distances are 62.226\,Mpc and 26.876\,Mpc for the maser and non-masing sources, respectively. This results in a distance bias of order (62.226\,/\,26.876)$^2$ $\sim$ 5.36. 
For these reasons, we eliminated all sources with D\,$>$ 70\,Mpc to acquire a minimum value of distance bias. This requires to reject 11 out of 35 maser sources, while none of the non-masing sources had to be taken out. The mean distances of the maser subsample with D\,$<$ 70\,Mpc and the sample devoid of detected masers become 36.874\,Mpc and 26.876\,Mpc, respectively, with a distance bias of order (36.874/26.876)$^2$ $\sim$ 1.88.

\begin{figure}
\centering
\includegraphics[width=\columnwidth]{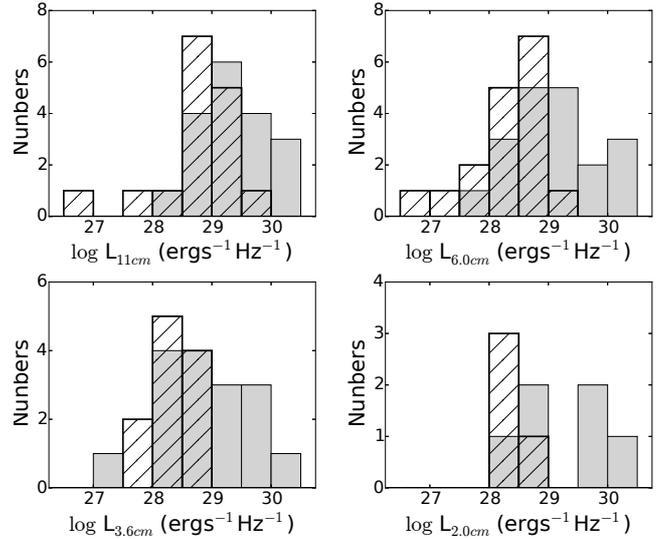}
\caption{Distributions of the 11cm (upper left panel), 6.0cm (upper right panel), 3.6cm (lower left panel) and 2.0cm (lower right panel) radio continuum luminosity for Seyfert 2s with distance D\,$\textless$\,70\,Mpc (grey: maser sources and open histograms with diagonal lines: sources without detected masers).}
\label{4distri_dis70}
\end{figure}

After accounting for the distances, there are small changes in mean values of luminosity for the maser subsample (for details, see Table.\ref{table3}), while there are no changes for the non-masing sample.
The differences between both samples in radio continuum luminosity become smaller but are still obvious, with luminosity ratios of 4.3\,$\pm$\,1.0, 5.0\,$\pm$\,1.2, 3.7\,$\pm$\,0.9 and 7.2\,$\pm$\,1.7 for $\emph L_{11cm}$, $\emph L_{6.0cm}$, $\emph L_{3.6cm}$ and $\emph L_{2.0cm}$, respectively.  

The distributions of $\log L_{11cm}$, $\log L_{6.0cm}$, $\log L_{3.6cm}$ and $\log L_{2.0cm}$ for the maser subsample (grey filled histograms) and non-masing sample (open histograms with diagonal lines) are plotted in Fig.\ref{4distri_dis70}. 
The logrank test results show that the difference of the radio continuum luminosity distribution at 11\,cm, 6.0\,cm and 3.6\,cm is still significant with a chance probability less than 0.05.
Within a similar distance range (D<\,70Mpc), the H$_2$O maser sources still have larger radio luminosity means than the non-masing ones, which is further supported by t-Test results. And considering the distance bias of order $\sim$\,1.88, the luminosity ratios become a factor of 2 to 3.

Over all, maser host Seyfert 2 galaxies have relatively higher radio continuum luminosities than those of non-masing Seyfert 2s which agrees with the proposition that the radio luminosity is a suitable indicator to guide future AGN maser searches \citep{2012A&A...538A.152Z}.

\subsection{Spectral properties}

\begin{figure}
\centering
\includegraphics[width=\columnwidth]{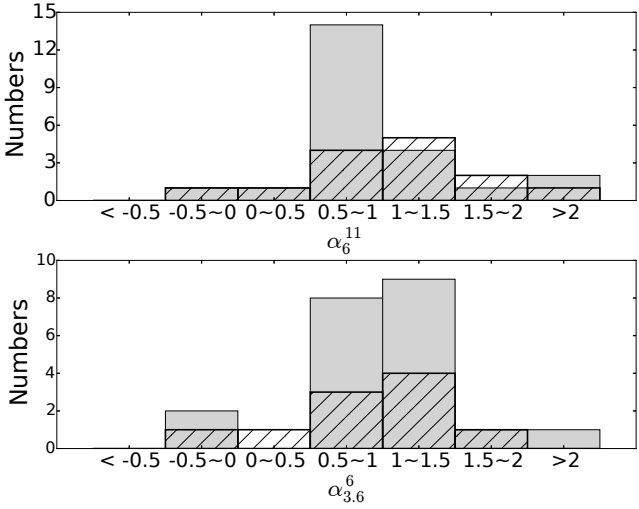}
\caption{The distributions of radio spectral indices of Seyfert2s (grey: maser sources; open histograms with diagonal lines: sources without detected masers). Upper panel: Spectral index between 11cm and 6cm ($\alpha_{6}^{11}$); lower panel: Spectral index between 6cm and 3.6cm ($\alpha_{3.6}^{6}$).}
\label{3si}
\end{figure}

Assuming a power-law dependence, which is given by S $\propto \lambda^{\alpha}$, the spectral index between two wavelengths $\lambda_1$ and $\lambda_2$ for a source can be calculated by: 
\begin{equation}
	\alpha_{\lambda 1}^{\lambda 2} = \log(\emph S_{\lambda_1}/\emph S_{\lambda_2})/\log(\lambda_1/\lambda_2)
\end{equation}
The spectral indices were calculated for both our maser and non-masing Seyfert 2 sources, of which flux densities of two adjacent bands were measured, and are listed in Table \ref{masergalaxies} (see columns 10 to 13), and in Table \ref{nonmaser} (see columns 8 to 10), respectively. Mean values are listed in Table \ref{table3}.

Here, we just made statistical comparisons on $\alpha_{6}^{11}$ and $\alpha_{3.6}^{6}$ for our samples, since the number of sources for $\alpha_{2}^{3.6}$ and $\alpha_{1.3}^{2}$ are too small (e.g., only four maser sources with effective $\alpha_{2}^{3.6}$).
For our maser Seyfert 2 sample, the mean values of $\alpha_{6}^{11}$ and $\alpha_{3.6}^{6}$ are 1.02\,$\pm$\,0.14 and 0.96\,$\pm$\,0.11, respectively.
And for our non-masing Seyfert 2 sample, the mean values of $\alpha_{6}^{11}$ and $\alpha_{3.6}^{6}$ are 1.01\,$\pm$\,0.16 and 1.01\,$\pm$\,0.17, respectively.
Comparisons show no significant differences in the mean values of both spectral indices between the maser and non-masing Seyfert 2 samples, which is further supported by t-Test results (listed in Table \ref{table3}).
Fig.\ref{3si} shows the histograms of $\alpha_{6}^{11}$ and $\alpha_{3.6}^{6}$ for both megamaser and non-masing samples. 
Within the histograms, similar distributions could be found between megamaser and non-masing sources for both spectral indices, which peaks around the 0.5-1.0 bin. This is further supported by our KS-Test results (see details in Table \ref{Lh2ovsL}), i.e., no significant difference on $\alpha_{6}^{11}$ and $\alpha_{3.6}^{6}$ between megamaser and non-masing Seyfert 2 galaxies. This is consistent with our previous study on $\alpha_{20}^{6}$ for both megamaser and non-masing samples \citep{2012A&A...538A.152Z}. There are slight changes in mean values of $\alpha_{6}^{11}$ and $\alpha_{3.6}^{6}$ after accounting for the distances, and comparisons only including sources at D\,$\textless$\,70\,Mpc give a very similar picture.

\begin{figure}
\centering 
\includegraphics[width=\columnwidth]{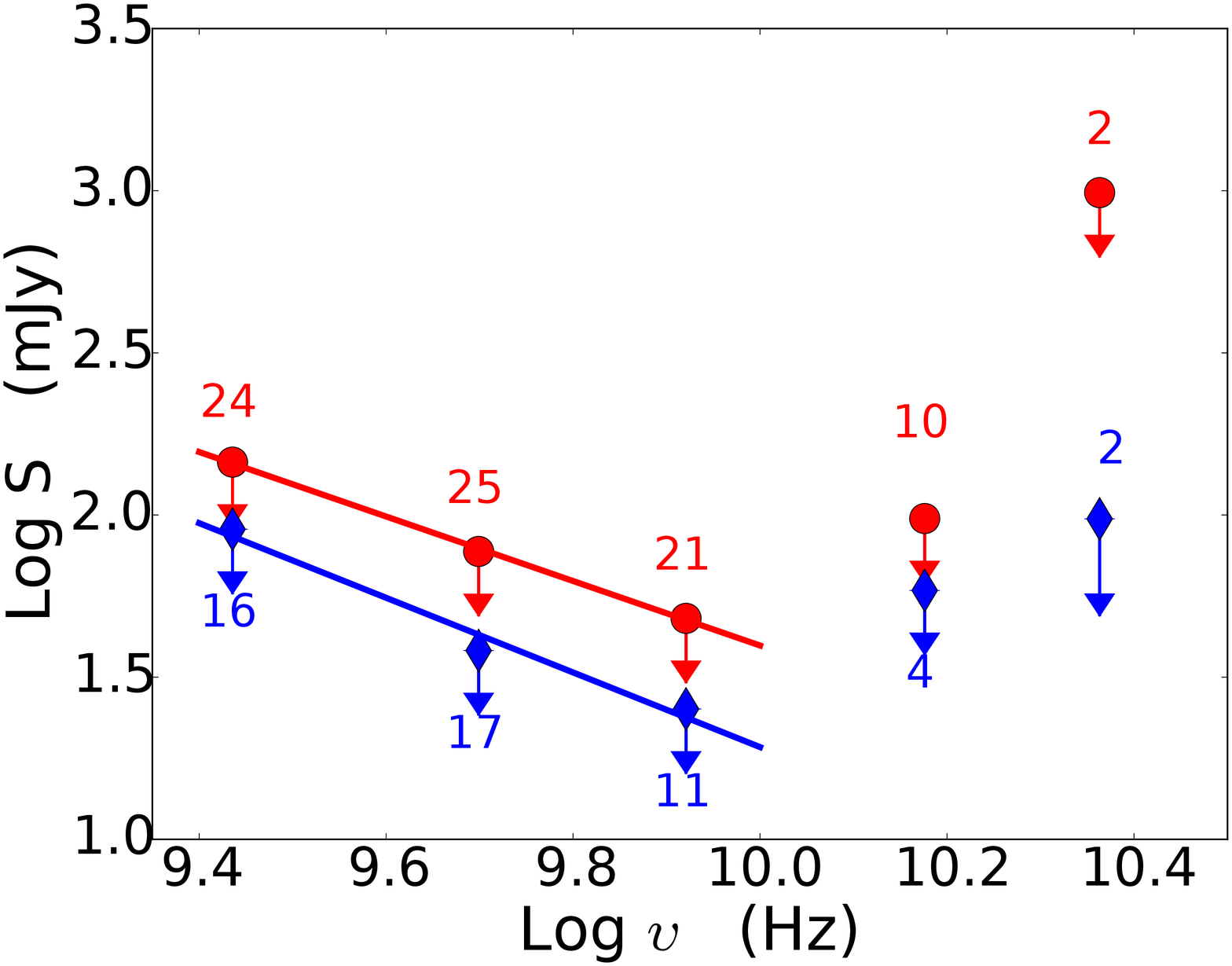}
\caption{The mean spectral energy distribution for the maser Seyfert 2 sample (red circles) and the non-masing Seyfert 2 sample (blue diamonds). The numbers of sources that are used to derive mean flux densities ($\log \rm S$) were added to each point in the diagram. Trend lines for mean flux densities at low frequencies (i.e. 11\,cm, 6.0\,cm and 3.6\,cm) are presented for both samples (red  line: maser Seyfert 2 sample; blue: non-masing Seyfert 2 sample).}
\label{sed}
\end{figure}

We have plotted the mean spectral energy distribution (SED) for our maser and non-masing Seyfert 2 samples in Fig.\ref{sed}. The numbers of sources that are used to derive mean flux densities ($\log \rm S$) are added to each point in the diagram. The mean flux densities are the upper limits of the whole sample in each frequency.
From the figure, the mean SED for both samples look similar. The same downtrend is apparent at low frequency bands (i.e. 11\,cm, 6.0\,cm and 3.6\,cm) for both Seyfert 2 samples. 
With slopes of -0.99\,$\pm$\,0.03 and -1.15\,$\pm$\,0.17 for the maser and the non-masing Seyfert 2 sample, respectively, the trend lines at the low frequency bands show that the mean flux density of Seyfert 2 galaxies decreases with frequency, which is consistent with the results of our previous work \citep{2012A&A...538A.152Z}. 
However, for both samples at high frequency (i.e., 2.0\,cm and 1.3\,cm), the mean flux density increases for both samples. The very low detection rate due to high noise levels in high frequency bands could account for the reversal in the mean SEDs. 
Observations at high frequency bands were subjected to a harsh selection effect, so that only a few bright sources could be detected with S/N larger than 3, e.g., for non-masing sources, only four sources were detected at 2\,cm and there was no detection at 1.3\,cm. This leads to the rather large mean flux density and ascendant trend in high frequency bands. More sensitive observations in the higher frequency bands are needed, especially for the K-band ($\lambda$ $\sim$ 1.3\,cm). 
Another striking feature in Fig.\ref{sed} is that the mean flux density of the maser Seyfert 2 sample is constantly larger than those of the non-masing ones at each band, roughly by a factor of 2. This is related to the difference in average luminosity, discussed in Sect.4.1.

\subsection{Radio continuum power versus H$_2$O megamaser power}

\begin{figure}
\centering
\includegraphics[width=\columnwidth]{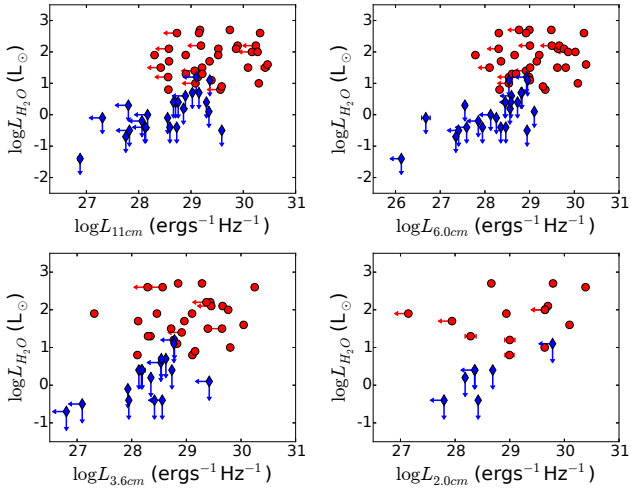}
\caption{H$_2$O maser luminosities (logarithmic scale in units of L$_\odot$) vs. radio luminosities (logarithmic scale in units of erg$\cdot$s$^{-1}$Hz$^{-1}$) of maser host galaxies (red circles) at 11\,cm (upper left panel), 6.0\,cm (upper right panel), 3.6\,cm (lower left panel) and 2.0\,cm (lower right panel), respectively. For comparison, estimated 5$\sigma$ upper limits of H$_2$O maser luminosity ($\log  L_{\rm H_2O}$, in units of L$_\odot$) of non-masing galaxies (blue diamonds) are also plotted.}
\label{h2ovslum}
\end{figure}


Since AGNs are considered to be the ultimate energy source for the H$_2$O megamaser emissions \citep{Lo:2005iq}, we may then expect to find some correlations between the power of the radio continuum and the H$_2$O megamaser emissions.
For our megamaser Seyfert 2 sample, the apparent luminosity of H$_2$O megamaser emission is plotted against $\log L_{11cm}$(upper left panel), $\log L_{6.0cm}$(upper right panel), $\log L_{3.6cm}$(lower left panel) and $\log L_{2.0cm}$(lower right panel) in Fig.\ref{h2ovslum}.
Assuming a linewidth of 20\,km\,s$^{-1}$, 5$\sigma$ upper limits of H$_2$O maser luminosity are derived from the individual rms values (taken from the MCP website, as mentioned in Sect.1) of H$_2$O maser data for the non-masing Seyfert 2s (e.g., \citealt{2009ApJ...695..276B}). For comparison, these upper limits (details in Table \ref{nonmaser}) are also plotted in Fig.\ref{h2ovslum}.
Owing to the low number of detections, the luminosity of the radio continuum in the 1.3\,cm band is not part of the following discussion.


From Fig.\ref{h2ovslum}, it is apparent that H$_2$O undetected sources tend to locate in the lower left region and maser sources in the upper right region, which is consistent with our previous results.
For the megamaser sample alone, there is no significant correlation between luminosities of the H$_2$O megamaser emission and the radio continuum, which is supported by Spearman's rank tests. The results listed in Table \ref{Lh2ovsL} indicate that there could be a positive but statistically weak correlation at each band. This is consistent with previous results of \citet{2012A&A...538A.152Z}. Adopting the scenario that the H$_2$O maser emission is mainly produced by amplification of the nuclear radio continuum emission, the lack of a correlation should be mainly caused by large uncertainties on both luminosities. 

\begin{table}
\caption[]{Statistical results on radio luminosity and H$_2$O maser luminosity for megamaser Seyfert 2s}
\label{Lh2ovsL}
\begin{tabular}{lcccc}
\hline
     				& 11\,cm				&  6.0\,cm	 			& 3.6\,cm 				& 2.0\,cm			\\ \hline
Spearman's Coeff.  	& 0.13 					& 0.06 					& 0.31 					& 0.24 			 	 	\\
Spearman's Prob.		& 0.54					& 0.77					& 0.18					& 0.51					\\ \hline 

\end{tabular}

\scriptsize
\textbf{Note:} the values are calculated from linear fits in four frequency bands. A positive coefficient indicates a positive correlation implying that radio luminosity and H$_2$O luminosity for megamaser Seyfert 2s tend to increases together. A correlation coefficient is significant if probability is less than the significance level of 0.05.
\end{table}

Uncertainties in H$_2$O maser luminosity possibly may arise from 
directional maser emission and different AGN related maser types \citep{2012A&A...538A.152Z}. 
For uncertainties in nuclear radio luminosity, two possible causes should be considered.
First, uncertainties may be introduced assuming the measured radio continuum luminosity as isotropic indicator of the intrinsic AGN power, since the observed radio power possibly depends on the AGN torus structure \citep{2012A&A...538A.152Z}. 
Another effect, which may be relevant here, is that the observed nuclear radio power could be contaminated by the large scale emission of the host galaxy \citep{2009ApJ...698..623D}. The beam size of the Effelsberg 100\,m radio telescope at our observed bands is at least 40", which corresponds to a linear size of $\sim$2\,kpc for a source at distance 10\,Mpc. 
This is much larger than that of the maser spots observed on sub-pc scales \citep[e.g.,][]{1995Natur.373..127M, 2009AGUFM.A14C..08R} and that of potentially associated nuclear continuum sources. Systematic interferometric observations with high resolution are required to resolve the scenario between nuclear continuum and maser spot distribution.

\subsection{Constraints on parameters of H$_2$O megamaser AGNs?}

Based on the fact that megamaser Seyfert 2s tend to have larger radio continuum luminosities (with robust limits of $L_\nu\geq$\,10$^{29}$\,erg$\cdot$s$^{-1}$Hz$^{-1}$, see Sect. 4.1), we try to investigate possible constraints on AGN parameters of H$_2$O megamaser hosts.
For both our Seyfert 2 samples, we compiled mean stellar velocity dispersions of 20 out of 35 megamaser sources and 24 out of 25 non-maser ones from HyperLeda (http://leda.univ-lyon1.fr/).
For an additional seven megamaser sources, their stellar velocity dispersions are taken from the literature (see details in Table \ref{para_maser}).
The corresponding BH masses for each source are calculated with the M\,-\,$\sigma$ relation
\begin{equation}
	\log(\frac{M_{\rm BH}}{M_{\odot}})=\alpha +\beta \log(\frac{\sigma}{200\, \rm km\,s^{-1}})\,,
\end{equation}
where $\alpha$\,=\,8.08 and $\beta$\,=\,4.47 \citep{2009ApJ...698..198G}. Here we have to mention, the M$_{\rm BH}$-Sigma relation defined by elliptical galaxies with high mass, may not provide good BH mass estimations for low-mass maser galaxy systems, which is proposed from precise BH mass measurements from disk megamaser sources \citep{2010ApJ...721...26G,Greene:2016wv}. For those common seven disk megamaser sources, our results from the M-Sigma relation are basically consistent with their results from the dynamics of disk megamasers, with the largest differences less than 0.8 dex(see details in Table \ref{para_maser}). Their measured BH mass results are used in our following analysis.


\begin{figure}
\centering
\includegraphics[width=\columnwidth]{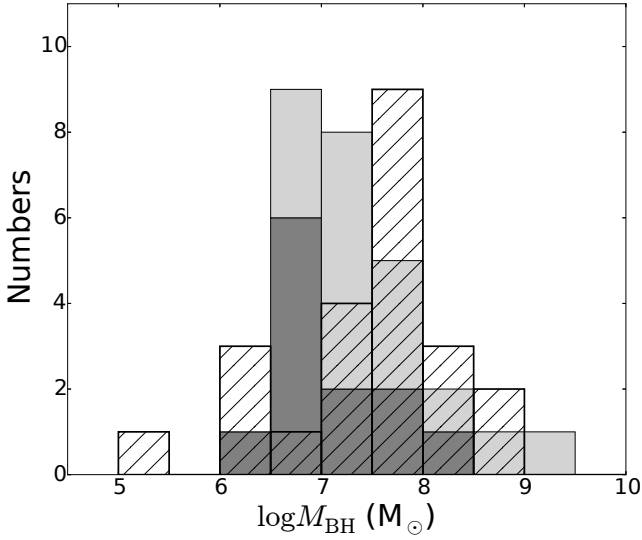}
\caption{The distributions of derived BH masses. Dark histograms: disk maser sources. Grey histograms: maser sources of other kinds. Open histograms with diagonal lines: non-masing galaxies.}
\label{figure9}
\end{figure}

Number distributions of derived BH masses for maser and non-masing samples are plotted in Fig.\ref{figure9}. The distributions of BH masses for both (also the disk maser subsample) samples mainly range from 10$^{6.5}M_\odot$ to $10^{8.5}M_\odot$ with a similar mean value of $\sim$\,10$^{7.35}M_\odot$. A logrank test gives a p-value of 0.54 suggesting that there is no significant difference on $M_{\rm BH}$ distributions for our maser Seyfert 2s and the non-masing Seyfert 2s, though non-masing Seyfert 2s tend to have larger peak values.
\begin{table}
\caption{Mean values of the dimensionless and mass accretion rates}
\label{paratable}
\resizebox{\columnwidth}{!}{%
\begin{tabular}{lccccc}
\hline
Sample      & Sub-sample			 & $\log \dot{m}_{11}$	 & $\log \dot{m}_6$		& $\log \dot{M}_{11}$ 	& $\log \dot{M}_6$			\\ \hline
Maser  		& Total				 & -6.61\,$\pm$\,0.13 & -6.64\,$\pm$\,0.14 	& -6.81\,$\pm$\,0.13 	& -6.86\,$\pm$\,0.14 		\\
			& Disk-maser			 & -6.64\,$\pm$\,0.20 &	-6.66\,$\pm$\,0.20	& -7.07\,$\pm$\,0.16	& -7.08\,$\pm$\,0.17\\
Non-masing  	& 	& -7.42\,$\pm$\,0.19 & -7.59\,$\pm$\,0.16 	& -7.55\,$\pm$\,0.16 	& -7.65\,$\pm$\,0.15 		\\ \hline
t-Test Prob.	& 	& 0.017 				& 0.001					& 0.001					& \textless 0.001		\\ \hline 

\end{tabular}
}
\scriptsize
\textbf{Note:} t-Test results for maser and non-masing samples.
\end{table}



Taken the radio continuum luminosity as  an isotropic luminosity indicator of AGN power \citep[e.g.,][]{2009ApJ...698..623D,1990ApJS...72..551G}, the dimensionless  accretion rate (i.e., Eddington ratio) can be estimated:
\begin{equation}
\dot{m}_{\nu}=\frac{L_\nu}{L_{\rm Edd}}\,,	
\end{equation}
where $L_{\nu}$ is the radio continuum luminosity and Eddington luminosities \emph L$_{\rm Edd}$ are calculated from 
\begin{equation}
	L_{\rm Edd}=1.3\times 10^{46}\left ( \frac{M_{\rm BH}}{10^{8}M_{\odot}} \right ) \rm erg\,s^{-1}\,.
\end{equation}
Adopting the standard accretion model, where within three times the Schwarzschild radius 3$r_{g} = 3\, \times \frac{2 \rm GM}{c^2}$, matter falls into the central black-hole and presumably half of the gravitational energy transforms into radiation, the mass accretion rate can be estimated by \cite{2001PASJ...53..215I}:
\begin{equation}
	\dot{M}_{\nu}=\frac{12L_\nu}{c^2}=2 \times 10^{-5} \left ( \frac{L_\nu}{1.0 \times 10^{41}\,\rm erg\,s^{-1}} \right ) M_\odot \rm \,yr^{-1}\,.
\end{equation}

Estimated values of the dimensionless $\dot{m}_{\nu}$ (Eq. 4) and mass accretion rates $\dot{M}_{\nu}$ (Eq. 6) are listed in Table \ref{para_maser} and \ref{para_nonmaser}, respectively. 

\begin{figure}
\centering
\includegraphics[width=\columnwidth]{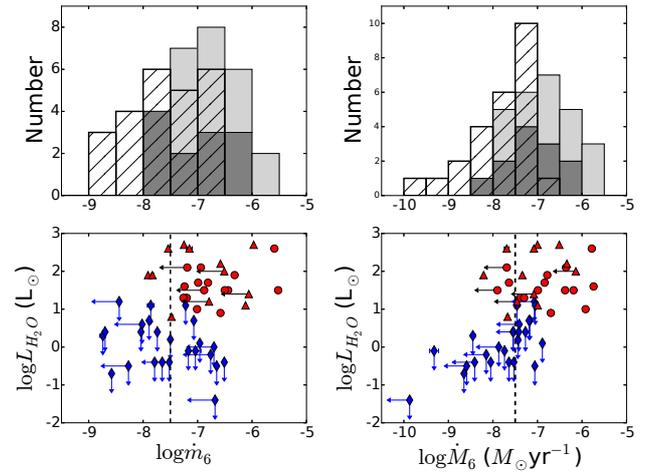}
\caption{Upper panels: Number distributions of logarithmic accretion rates, specifically referring to $\log \dot{m_6}$ (left), defined by Eq.(4), and $\log \dot{M_6}$ (right), defined by Eq.(6). The indices indicate that the data are derived from radio continuum luminosities at 6.0\,cm. Dark histograms: disk maser sources. Grey histograms: maser sources of other kinds. Open histrograms with diagonal lines: non-masing galaxies. Lower panels: H$_2$O megamaser luminosities vs. $\log \dot{m_6}$ (left) and $\log \dot{M_6}$ (right). Circles and triangles indicate possible disk maser sources and maser sources of other kinds, respectively. Blue diamonds are non-maser sources. Dashed lines: $\log \dot{m_6}$ = -7.5 and $\log \dot{M_6}$ = -7.5.}
\label{figure10}
\end{figure}

Fig.\ref{figure10} (upper panels) presents the distributions of  accretion rates for both maser and non-masing samples. It shows that the maser Seyfert 2s tend to have higher accretion rates than the non-masing ones. The mean accretion rates of maser Seyfert 2s are nearly one order of magnitude larger than those of non-masing Seyfert 2s (Table \ref{paratable}). No significant difference can be found between disk-masers and other types of AGN masers.

In the lower panels of Fig.\ref{figure10}, maser luminosities are plotted against both accretion rates that are derived from radio continuum luminosities at 6.0\,cm (in logarithmic scale, $\log \dot{m_6}$ and $\log \dot{M_6}$). The difference in accretion rate between maser and non-masing samples is obvious, i.e., maser Seyfert 2s tend to possess larger accretion rates. For most of maser Seyfert 2s, the dimensionless (Eq. 4) and mass accretion (Eq. 6) rate are larger than 10$^{-7.5}$ and 10$^{-7.5}$ $M_\odot\rm yr^{-1}$, respectively (dotted lines in the lower panels of Fig.\ref{figure10}). These agree with the limited criteria of $L_\nu\geq$\,10$^{29}$\,erg$\cdot$s$^{-1}$Hz$^{-1}$ we proposed in Sect. 4.1, with a mean BH mass of 10$^8\,M_{\odot}$.

\section{Summary}

In this paper, multi-band radio continuum observations from the Effelsberg 100\,m telescope are presented targeting H$_2$O megamaser host Seyfert 2s and a control Seyfert 2 sample without maser detection.
Radio properties of these two samples were compared to obtain a better understanding of intrinsic radio properties of H$_2$O maser host galaxies, the formation of such megamasers, and to provide a better guidance for future megamaser surveys.

Megamaser Seyfert 2 galaxies tend to possess larger radio luminosity than Seyfert 2s without maser detection.
The mean radio continuum luminosities for maser Seyfert 2s that are detected are $\log L_{11cm}$ = 29.49\,$\pm$\,0.02, $\log L_{6.0cm}$ = 29.18\,$\pm$\,0.01, $\log L_{3.6cm}$ = 29.02\,$\pm$\,0.02, $\log L_{2.0cm}$ = 29.35\,$\pm$\,0.04 and $\log L_{1.3cm}$ = 30.09\,$\pm$\,0.03, and those of non-masing Seyfert 2s are $\log L_{11cm}$ = 28.71\,$\pm$\,0.02, $\log L_{6.0cm}$ = 28.35\,$\pm$\,0.02, $\log L_{3.6cm}$ = 28.35\,$\pm$\,0.03 and $\log L_{2.0cm}$ = 28.41\,$\pm$\,0.03, respectively.
Considering a distance bias of order $\sim$\,1.88, the luminosity difference remains significant, with luminosity ratios of order 2-3.

For both samples, spectral indices are derived between two adjacent bands.
The mean values of $\alpha_{11cm}^{6cm}$ and $\alpha_{6cm}^{3.6cm}$ are 1.02\,$\pm$\,0.14, 0.96\,$\pm$\,0.11 for the maser Seyfert 2 sample and 1.01\,$\pm$\,0.16, 1.01 $\pm$ 0.17 for the non-masing Seyfert 2 samples, respectively. Comparisons on distributions of spectral indices show no significant differences.  
Due to large uncertainties in the H$_2$O isotropic luminosity and radio luminosity of maser host Seyfert 2s, the statistical correlation is not obvious between them. 

Overall, our results confirm the trend that H$_2$O maser host Seyfert 2 galaxies have larger radio luminosity than non-masing Seyfert 2s. Taking the radio luminosity as an isotropic tracer of AGN power, thus megamaser Seyfert 2s have stronger AGN power than non-masing Seyfert 2s. The black hole mass, the dimensionless and mass accretion rate were estimated for our maser Seyfert 2s and non-masing Seyfert 2s. It shows that the accretion rates of maser Seyfert 2s are nearly one order larger than non-masing Seyfert 2s. This may provide possible connection between H$_2$O megamaser formation and AGN activity, as well suitable constraints on future megamaser surveys.

\section*{Acknowledgements}
 This work is supported by China Ministry of Science and Technology under State Key Development Program for Basic Research (2012CB821800) and the Natural Science Foundation of China (No. 11473007, 11590782, 11590780, 11590784).

\newpage
\begin{table*}
\caption{Radio properties of H$_2$O megamaser Seyfert 2 galaxies}
\resizebox{\textwidth}{!}{%
\begin{tabular}{lcccccccccccc}
\hline 
Source          & Type  & D     & $\log$ \emph L$_{\rm H_2O}$ & \emph S$_{11}$      & \emph S$_{6}$       & \emph S$_{3.6}$     & \emph S$_{2}$       & \emph S$_{1.3}$     & $\alpha_{6}^{11}$ & $\alpha_{3.6}^{6}$ & $\alpha_{2}^{3.6}$ & $\alpha_{1.3}^{2}$ \\ \hline
\textbf{NGC 17}          & Sy2   & 79.1         & 0.8      & \textless 48.85     & 30.03 $\pm$ 1.06    & 17.25 $\pm$ 0.95    & 13.24 $\pm$ 2.51    & -                   & -     & 1.09  & 0.45  & -     \\
Mrk 348         & Sy2   & 62           & 2.6      & 455.89 $\pm$ 6.56   & 358.54 $\pm$ 4.65   & 388.64 $\pm$ 8.05   & 531.92 $\pm$ 13.87  & 723.42 $\pm$ 48.98  & 0.4   & -0.16 & -0.53 & -0.71 \\
Mrk 1           & Sy2   & 25           & 1.7      & 50.86 $\pm$ 2.89    & 30.69 $\pm$ 1.48    & 17.55 $\pm$ 1.13    & \textless 11.74     & -                   & 0.83  & 1.09  & -     & -     \\
\textbf{NGC 591}         & Sy2   & 61           & 1.4      & \textless 26.48     & \textless 18.69     & \textless 18.48     & -                   & -                   & -     & -     & -     & -     \\
IC 0184         & Sy2   & 70.5         & 1.0      & \textless 20.48     & \textless 5.78      & -                   & -                   & -                   & -     & -     & -     & -     \\
NGC 1052        & Sy2   & 17           & 2.1      & 965.5 $\pm$ 7.75    & 1253.17 $\pm$ 14.64 & 1334.14 $\pm$ 26.55 & 1444.74 $\pm$ 31.49 & 1345.92 $\pm$ 74.55 & -0.43 & -0.12 & -0.14 & 0.16  \\
\textbf{NGC 1068}        & Sy2   & 14.5         & 2.2      & 3100.99 $\pm$ 25.47 & 1830.81 $\pm$ 20.95 & 1092.31 $\pm$ 22.04 & -           & -                   & 0.87  & 1.01  & -     & -     \\
NGC 1106        & Sy2   & 57.8         & 0.9      & 96.54 $\pm$ 2.69    & 49.92 $\pm$ 1.67    & 36.19 $\pm$ 1.47    & -                   & -                   & 1.09  & 0.63  & -     & -     \\
Mrk 1066        & Sy2   & 48           & 1.5      & 59.24 $\pm$ 2.63    & 37.34 $\pm$ 1.47    & 19.14 $\pm$ 1.76    & -                   & -                   & 0.76  & 1.31  & -     & -     \\
\textbf{NGC 1320}        & Sy2   & 35.5         & 1.2      & \textless 24.26     & \textless 23.04     & -                   & -                   & -                   & -     & -     & -     & -     \\
IRAS 0335+0104  & Sy2   & 159.1        & 2.1      & 24.05 $\pm$ 1.37    & \textless 19.35     & \textless 9.44      & -                   & -                   & -     & -     & -     & -     \\
UGC 3255        & Sy2   & 75           & 1.2      & 20.18 $\pm$ 1.89    & 14.45 $\pm$ 0.96    & 8.77 $\pm$ 0.69     & 14.91 $\pm$ 3.20     & -                   & 0.55  & 0.98  & -0.9  & -     \\
Mrk 3           & Sy2   & 54           & 1.0      & 568.04 $\pm$ 8.74   & 345.02 $\pm$ 4.27   & 182.56 $\pm$ 3.94   & 125.25 $\pm$ 6.81   & \textless 170.28    & 0.82  & 1.25  & 0.64  & -     \\
VII Zw 073      & Sy2   & 158.9        & 2.2      & \textless 57.63     & \textless 10.41     & \textless 7.71      & -                   & -                   & -     & -     & -     & -     \\
\textbf{NGC 2273}        & Sy2   & 24.5         & 0.8      & 51.59 $\pm$ 3.62    & 29.64 $\pm$ 1.12    & 17.72 $\pm$ 0.86    & -                   & -                   & 0.91  & 1.01  & -     & -     \\
\textbf{UGC 3789}        & Sy2   & 44.3         & 2.6      & \textless 22.99     & 8.73 $\pm$ 1.01     & \textless 8.32      & -                   & -                   & -     & -     & -     & -     \\
Mrk 78          & Sy2   & 150          & 1.5      & 97.51 $\pm$ 17.57   & 15.58 $\pm$ 0.94    & 9.2 $\pm$ 0.79      & -                   & -                   & 3.03  & 1.03  & -     & -     \\
Mrk 1210        & Sy2   & 54           & 1.9      & 95.11 $\pm$ 2.19    & 47.19 $\pm$ 1.41    & 36.52 $\pm$ 1.08    & 24.89 $\pm$ 2.28    & -                   & 1.16  & 0.5   & 0.65  & -     \\
NGC 2979        & Sy2   & 36           & 2.1      & \textless 24.41     & \textless 13.09     & -                   & -                   & -                   & -     & -     & -     & -     \\
\textbf{NGC 3079}        & Sy2   & 15.5         & 2.7      & 517.99 $\pm$ 4.68   & 344.51 $\pm$ 4.11   & 246.19 $\pm$ 4.95   & 160.33 $\pm$ 4.11   & \textless 221.44    & 0.67  & 0.66  & 0.73  & -     \\
\textbf{Mrk 34}          & Sy2   & 205          & 2.0      & \textless 29.14     & \textless 14.50      & -                   & -                   & -                   & -     & -     & -     & -     \\
\textbf{NGC 3393}        & Sy2   & 50           & 2.6      & 47.71 $\pm$ 2.43    & 28.01 $\pm$ 1.52    & \textless 12.3      & -                   & -                   & 0.88  & -     & -     & -     \\
NGC 3735        & Sy2   & 36           & 1.3      & 106.17 $\pm$ 3.81   & 26.22 $\pm$ 0.99    & 14.24 $\pm$ 0.89    & -                   & -                   & 2.31  & 1.2   & -     & -     \\
\textbf{NGC 4258}        & Sy1.9 & 7.2          & 1.9      & 324.67 $\pm$ 4.47   & 98.89 $\pm$ 2.48    & 33.29 $\pm$ 0.97    & \textless 22.49     & -                   & 1.96  & 2.13  & -     & -     \\
\textbf{NGC 4388}        & Sy2   & 34           & 1.1      & 177.51 $\pm$ 1.95   & 73.18 $\pm$ 1.27    & 48.92 $\pm$ 1.53    & -                   & -                   & 1.46  & 0.79  & -     & -     \\
NGC 5256        & Sy2   & 112          & 1.5      & 82.14 $\pm$ 1.86    & 42.89 $\pm$ 1.19    & \textless 30.01                   & -                   & -                   & 1.07  & -     & -     & -     \\
NGC 5347        & Sy2   & 31           & 1.5      & \textless 23.02     & \textless 11.08     & -                   & -                   & -                   & -     & -     & -     & -     \\
NGC 5506        & Sy1.9 & 25           & 1.7      & 303.26 $\pm$ 3.54   & 193.7 $\pm$ 2.73    & 123.16 $\pm$ 2.64   & -                   & \textless 149.03    & 0.74  & 0.89  & -     & -     \\
\textbf{NGC 5728}        & Sy2   & 37           & 1.9      & 48.4 $\pm$ 2.17     & 28.25 $\pm$ 1.13    & \textless 17.68     & -                   & -                   & 0.89  & -     & -     & -     \\
\textbf{NGC 5793}        & Sy2   & 47           & 2.0      & 709.73 $\pm$ 7.48   & 399.52 $\pm$ 5.03   & 220.84 $\pm$ 4.77   & \textless 167.47                   & -                   & 0.95  & 1.16  & -     & -     \\
NGC 6240        & Sy2   & 98           & 1.6      & 253.21 $\pm$ 3.85   & 159.61 $\pm$ 2.66   & 97.45 $\pm$ 2.64    & 108.36 $\pm$ 5.19   & \textless 144.44    & 0.76  & 0.97  & -0.18 & -     \\
\textbf{NGC 6323}        & Sy2   & 104          & 2.7      & -                   & \textless 4.50       & -                   & -                   & -                   & -     & -     & -     & -     \\
IRAS F1937-0131 & Sy2   & 80           & 2.2      & \textless 20.29     & \textless 13.31     & -                   & -                   & -                   & -     & -     & -     & -     \\
\textbf{NGC 6926}        & Sy2   & 80           & 2.7      & 72.41 $\pm$ 2.42    & 40.02 $\pm$ 1.08    & 25.31 $\pm$ 1.47    & 81.06 $\pm$ 4.92    & -                   & 0.98  & 0.90   & -1.98 & -     \\
NGC 7479        & Sy2   & 31.8         & 1.3      & 66.24 $\pm$ 2.52    & 41.17 $\pm$ 1.68    & 16.46 $\pm$ 1.40     & 15.90 $\pm$ 3.67     & \textless 165.51    & 0.78  & 1.79  & 0.06  & -    \\ \hline
\end{tabular}
}
\label{masergalaxies}
\begin{flushleft}
	\scriptsize
	\textbf{Note:} Column (1): Source name (15 out of 35 are possible disk-masers in bold text). Column (2): Types of nuclear activity from \citet{2010ApJ...708.1528Z}. Column (3): Luminosity distance in units of Mpc, assuming H$_0$ = 70\,kms$^{-1}$Mpc$^{-1}$. NGC 17 and NGC 1320 from \citet{2008ApJ...686L..13G}. Column (4): Apparent luminosity of maser emission (on a logarithmic scale), in units of L$_\odot$, taken from \citet{2011A&A...532A.125T}, \citet{2009ApJ...695..276B}, \citet{2008ApJ...686L..13G}, and \citet{2008ApJ...685L..39D}. Columns (5) to (9): Observed flux densities at 11\,cm, 6.0\,cm, 3.6\,cm, 2.0\,cm and 1.3\,cm, in mJy, respectively. Column (10) to (13): Spectral indices assuming S $\propto \nu^{-\alpha}$.
\end{flushleft}
\end{table*}

\begin{table*}
\caption{Radio properties of Seyfert 2 galaxies without H$_2$O maser detections}
\resizebox{\textwidth}{!}{%
\begin{tabular}{lcccccccccccc}
\hline
Source   & Type  & D     & rms-H$_2$O & UL-H$_2$O & \emph S$_{11}$      & \emph S$_{6}$     & \emph S$_{3.6}$   & \emph S$_{2}$      & \emph S$_{1.3}$    & $\alpha_{6}^{11}$ & $\alpha_{3.6}^{6}$ & $\alpha_{2}^{3.6}$ \\ \hline
NGC 777  & Sy2   & 66.5 & 6.9 & 1.2  & \textless 23.86     & \textless 16.14   & \textless 11.46   & -                 & -                & -     & -     & -     \\
NGC 1058 & Sy2   & 9.2  & 20  & -0.1 & \textless 20.03     & 4.71 $\pm$ 1.04   & -                 & -                 & -                & -     & -     & -     \\
NGC 1358 & Sy2   & 52.6 & 3   & 0.6  & \textless 23.64     & \textless 11.88   & \textless 10.35   & -                 & -                & -     & -     & -     \\
NGC 1667 & Sy2   & 61.2 & 6.6 & 1.1  & 51.11 $\pm$ 2.8     & 20.08 $\pm$ 1.44  & 13.24 $\pm$ 0.98  & \textless 135.91  & -                & 1.54  & 0.82  & -     \\
NGC 2655 & Sy2   & 24.4 & 5.2 & 0.2  & 100.98 $\pm$ 7.78   & 51.04 $\pm$ 1.46  & 31.54 $\pm$ 2.17  & 21.46 $\pm$ 1.73  & -                & 1.13  & 0.94  & 0.66  \\
NGC 2992 & Sy1.9 & 34.1 & 2.3 & 0.1  & 158.54 $\pm$ 3.09   & 91.68 $\pm$ 1.81  & \textless 188.82  & -                 & -                & 0.9   & -     & -     \\
NGC 3147 & Sy2   & 40.9 & 6.3 & 0.7  & 53.04 $\pm$ 1.74    & 33.73 $\pm$ 1.11  & 21.11 $\pm$ 1.55  & -                 & -                & 0.75  & 0.92  & -     \\
NGC 3185 & Sy2   & 21.3 & 3   & -0.2 & \textless 21.31     & \textless 13.12   & -                 & -                 & -                & -     & -     & -     \\
NGC 3976 & Sy2   & 37.7 & 16  & 1.1  & -                   & 20.32 $\pm$ 2.9   & -                 & -                 & -                & -     & -     & -     \\
NGC 3982 & Sy1.9 & 17   & 3   & -0.4 & 37.93 $\pm$ 2.01    & 25.54 $\pm$ 1.76  & \textless 104.74  & \textless 18.15   & -                & 0.65  & -     & -     \\
NGC 4138 & Sy1.9 & 17   & 2.4 & -0.5 & 1122.88 $\pm$ 42.71 & 254.58 $\pm$ 3.38 & -                 & -                 & -                & 2.45  & -     & -     \\
NGC 4168 & Sy1.9 & 16.8 & 15  & 0.3  & \textless 18.83     & 10.58 $\pm$ 1.12  & -                 & -                 & -                & -     & -     & -     \\
NGC 4395 & Sy1.8 & 4.6  & 3.8 & -1.4 & 29.97 $\pm$ 3.18    & \textless 5.38                 & -                 & -                 & -                & -     & -     & -     \\
NGC 4472 & Sy2   & 16.8 & 17  & 0.4  & 150.3 $\pm$ 2.68    & 83.96 $\pm$ 1.72  & 44.6 $\pm$ 1.56   & 68.34 $\pm$ 3.3   & -                & 0.96  & 1.24  & -0.73 \\
NGC 4501 & Sy2   & 16.8 & 3   & -0.4 & 158.19 $\pm$ 2.25   & 67.31 $\pm$ 1.24  & 26.22 $\pm$ 1.47  & -                 & -                & 1.41  & 1.85  & -     \\
NGC 4565 & Sy1.9 & 9.7  & 4.6 & -0.7 & 50.41 $\pm$ 1.72    & 19.94 $\pm$ 0.97  & \textless 5.66    & -                 & -                & 1.53  & -     & -     \\
NGC 4579 & Sy1.9 & 16.8 & 3   & -0.4 & 117.08 $\pm$ 2.75   & 87 $\pm$ 1.38     & 77.37 $\pm$ 1.82  & 78.32 $\pm$ 3.59  & \textless 145.27 & 0.49  & 0.23  & -0.02 \\
NGC 4594 & Sy1.9 & 20   & 13  & 0.4  & 96.65 $\pm$ 2.14    & 112.34 $\pm$ 1.78 & 113.87 $\pm$ 3.29 & 102.27 $\pm$ 6.58 & -                & -0.25 & -0.03 & 0.18  \\
NGC 4725 & Sy2   & 12.4 & 4.1 & -0.5 & \textless 36.2      & \textless 14.03   & \textless 6.76    & -                 & -                & -     & -     & -     \\
NGC 4941 & Sy2   & 16.8 & 3   & -0.4 & \textless 40.26     & \textless 11.58   & -                 & -                 & -                & -     & -     & -     \\
NGC 5005 & Sy2   & 21.3 & 10  & 0.4  & 104.96 $\pm$ 2.53   & 52.97 $\pm$ 1.48  & 24.93 $\pm$ 1.17  & -                 & \textless 65.15  & 1.13  & 1.48  & -     \\
NGC 5395 & Sy2   & 46.7 & 4.1 & 0.7  & 54.46 $\pm$ 1.59    & 24.81 $\pm$ 1.07  & 13.36 $\pm$ 1.28  & -                 & -                & 1.3   & 1.21  & -     \\
NGC 5899 & Sy2   & 42.8 & 3   & 0.4  & 90.18 $\pm$ 4.22    & \textless 17.43   & 7.12 $\pm$ 0.91   & \textless 10.39   & -                & -     & -     & -     \\
NGC 6951 & Sy2   & 24.1 & 3   & -0.1 & 51.42 $\pm$ 2.18    & 25.98 $\pm$ 1.01  & 12.25 $\pm$ 0.86  & -                 & -                & 1.13  & 1.47  & -     \\
NGC 7743 & Sy2   & 24.4 & 3   & 0.0  & \textless 20.80      & \textless 18.92   & -                 & -                 & -                & -     & -     & -   \\ \hline

\end{tabular}

}
\label{nonmaser}
\begin{flushleft}
	\scriptsize
	\textbf{Note:} Column (1): Source name. Column (2): Optical classification from \citet{1995ApJ...454...95M} or \citet{1997ApJS..112..315H}. Column (3): Luminosity distance in units of Mpc, taken from \citet{2009ApJ...698..623D}. Column (4): Rms values of H$_2$O maser data in units of mJy for a 20\,kms$^{-1}$ wide channel, wich were taken from MPC and HoME wabpages. Column(5): Estimated 5$\sigma$ upper limits of H$_2$O maser luminosity ($\log$ L$_{H_2O}$, in units of L$_\odot$) for non-masing Seyfert 2 galaxies from rms values. Columns (6) to (10): Observed flux densities at 11\,cm, 6.0\,cm, 3.6\,cm and 2.0\,cm, respectively. Columns (11) to (13): Spectral indice assuming S $\propto \nu^{-\alpha}$.
\end{flushleft}

\end{table*}

\begin{table*}
\caption[]{The Parameters of 27 (out of 35) galaxies of the Megamaser Seyfert 2 sample}
\label{para_maser}
\resizebox{1\textwidth}{!}{%
\begin{tabular}{lcccccccc}
\hline
Source            & $\sigma$ & $\log$ M$_{\rm BH}$ & $\log$ \emph L$_{\rm H_2O}$ & $\log$ \emph L$_{\rm Edd}$ & $\log \dot{m}_{11}$        & $\log \dot{m}_{6.0}$       & $\log \dot{M}_{11}$                      & $\log \dot{M}_{6.0}$                     \\ \hline
Mrk 348           & 141 & 7.40 & 2.6 & 45.51 & -5.75$\pm$0.01  & -5.59$\pm$0.01  & -5.94 $\pm$ 0.01 & -5.78 $\pm$ 0.01 \\
Mrk 1             & 111 & 6.94 & 1.7 & 45.05 & -7.03$\pm$0.02  & -6.99$\pm$0.02  & -7.68 $\pm$ 0.02 & -7.64 $\pm$ 0.02 \\
NGC 1052          & 210 & 8.17 & 2.1 & 46.28 & -7.32$\pm$0.01  & -6.94$\pm$0.01  & -6.74 $\pm$ 0.01 & -6.36 $\pm$ 0.01 \\
NGC 1106          & 146 & 7.47 & 0.9 & 45.58 & -6.56$\pm$0.01  & -6.58$\pm$0.01  & -6.68 $\pm$ 0.01 & -6.70 $\pm$ 0.01 \\
Mrk 1066          & 117 & 7.04 & 1.5 & 45.15 & -6.50$\pm$0.02  & -6.44$\pm$0.02  & -7.05 $\pm$ 0.02 & -6.99 $\pm$ 0.02 \\
Mrk 3             & 273 & 8.68 & 1.0 & 46.79 & -7.06$\pm$0.01  & -7.01$\pm$0.01  & -5.97 $\pm$ 0.01 & -5.92 $\pm$ 0.01 \\
Mrk 78            & 165 & 7.71 & 1.5 & 45.82 & -5.97$\pm$0.08  & -6.50$\pm$0.03  & -5.84 $\pm$ 0.08 & -6.38 $\pm$ 0.03 \\
Mrk 1210          & 114 & 7.12 & 1.9 & 45.23 & -6.27$\pm$0.01  & -6.32$\pm$0.01  & -6.74 $\pm$ 0.01 & -6.78 $\pm$ 0.01 \\
NGC 2979          & 112 & 7.09 & 2.1 & 45.20 & \textless -7.19 & \textless -7.19 & \textless -7.69  & \textless -7.69  \\
NGC 3735          & 141 & 7.40 & 1.3 & 45.51 & -6.86$\pm$0.02  & -7.20$\pm$0.02  & -7.05 $\pm$ 0.02 & -7.39 $\pm$ 0.02 \\
NGC 5256          & 100 & 6.92 & 1.5 & 45.03 & -5.50$\pm$0.01  & -5.52$\pm$0.01  & -6.17 $\pm$ 0.01 & -6.19 $\pm$ 0.01 \\
NGC 5347          & 92  & 6.57 & 1.5 & 44.68 & \textless -6.82 & \textless -6.88 & \textless -7.84  & \textless -7.9   \\
NGC 5506          & 180 & 7.56 & 1.7 & 45.67 & -6.88$\pm$0.01  & -6.81$\pm$0.01  & -6.91 $\pm$ 0.01 & -6.84 $\pm$ 0.01 \\
NGC 6240          & 336 & 9.09 & 1.6 & 47.20 & -7.30$\pm$0.01  & -7.24$\pm$0.01  & -5.80 $\pm$ 0.01 & -5.74 $\pm$ 0.01 \\
NGC 7479          & 152 & 7.55 & 1.3 & 45.66 & -7.32$\pm$0.02  & -7.26$\pm$0.02  & -7.36 $\pm$ 0.02 & -7.30 $\pm$ 0.02 \\ 
\textbf{Mrk 34}   & 181 & 7.96 & 2.0 & 46.07 & \textless -6.47 & \textless -6.51 & \textless -6.10  & \textless -6.14  \\
\textbf{NGC 5728} & 200 & 8.08 & 1.9 & 46.19 & -7.86$\pm$0.02  & -7.83$\pm$0.02  & -7.36 $\pm$ 0.02 & -7.34 $\pm$ 0.02 \\
\textbf{NGC 6926} & 109 & 7.04 & 2.7 & 45.15 & -5.97$\pm$0.01  & -5.97$\pm$0.01  & -6.52 $\pm$ 0.01 & -6.51 $\pm$ 0.01 \\
\textbf{NGC 591}  & 92  & 6.57 & 1.4 & 44.68 & \textless -6.17 & \textless -6.06 & \textless -7.19  & \textless -7.08  \\
\hline
\textbf{UGC 3789} & 107 & 7.05 & 2.6 & 45.16 & \textless -6.99 & -7.15$\pm$0.05  & \textless -7.53  & -7.69 $\pm$ 0.05 \\
\textbf{UGC 3789$^{(\ast)}$} &  & 6.99 & 2.6 & 45.10 & \textless -6.93 & -7.09$\pm$0.05  &  &  \\
\textbf{NGC 1068} & 176 & 7.83 & 2.2 & 45.94 & -6.61$\pm$0.01  & -6.58$\pm$0.01  & -6.37 $\pm$ 0.01 & -6.34 $\pm$ 0.01 \\
\textbf{NGC 1068$^{(\ast)}$} & 	& 6.92 & 2.2 & 45.03 & -5.70$\pm$0.01  & -5.67$\pm$0.01  & & \\
\textbf{NGC 1320} & 110 & 6.92 & 1.2 & 45.03 & \textless -7.03 & \textless -6.79 & \textless -7.70  & \textless -7.46  \\
\textbf{NGC 1320$^{(\ast)}$} & & 6.74 & 1.2 & 44.85 & \textless -6.85 & \textless -6.61 & &   \\
\textbf{NGC 2273} & 141 & 7.40 & 0.8 & 45.51 & -7.51$\pm$0.03  & -7.48$\pm$0.02  & -7.70 $\pm$ 0.03 & -7.67 $\pm$ 0.02 \\
\textbf{NGC 2273$^{(\ast)}$} & & 6.93 & 0.8 & 45.04 & -7.04$\pm$0.03  & -7.01$\pm$0.02  & & \\
\textbf{NGC 3079} & 176 & 7.83 & 2.7 & 45.94 & -7.33$\pm$0.01  & -7.25$\pm$0.01  & -7.09 $\pm$ 0.01 & -7.00 $\pm$ 0.01 \\
\textbf{NGC 3079$^{(\ast)}$} & & 6.4 & 2.7 & 44.51 & -5.9$\pm$0.01  & -5.82$\pm$0.01  & &  \\
\textbf{NGC 3393} & 197 & 8.05 & 2.6 & 46.16 & -7.57$\pm$0.02  & -7.54$\pm$0.02  & -7.11 $\pm$ 0.02 & -7.08 $\pm$ 0.02 \\
\textbf{NGC 3393$^{(\ast)}$} & & 7.2 & 2.6 & 43.31 & -6.72$\pm$0.02  & -6.69$\pm$0.02  & & \\
\textbf{NGC 4258} & 133 & 7.29 & 1.9 & 45.40 & -7.66$\pm$0.01  & -7.91$\pm$0.01  & -7.96 $\pm$ 0.01 & -8.21 $\pm$ 0.01 \\
\textbf{NGC 4258$^{(\ast)}$} & & 7.58 & 1.9 & 45.69 & -7.95$\pm$0.01  & -8.20$\pm$0.01  & & \\
\textbf{NGC 4388} & 99  & 6.71 & 1.1 & 44.82 & -5.99$\pm$0.01  & -6.12$\pm$0.01  & -6.87 $\pm$ 0.01 & -7.00 $\pm$ 0.0  \\
\textbf{NGC 4388$^{(\ast)}$} & & 6.86 & 1.1 & 44.79 & -5.96$\pm$0.01  & -6.09$\pm$0.01  & &  \\

\hline
\end{tabular}
}
\begin{flushleft}
	\scriptsize
	\textbf{Note:} 
Column (1): Source name (12 possible disk-masers are in bold text). 
Column (2): The average stellar velocity dispersion (in units of $km\,s^{-1}$) from HypeLeda (http://leda.univ-lyon1.fr/). Additionally, the average stellar velocity dispersion for UGC3789 is taken from \cite{Greene:2016wv} and for Mrk1210, NGC2979, Mrk34, NGC5256, NGC5056 \& NGC6926 are taken from \cite{2008ChJAA...8..547S} and \cite{wangjin2010}.
Column (3): Blackhole mass, in units of M$_\odot$, derived from the empirical M-$\sigma$ relation. For comparison, the dynamical BH mass value for those seven disk-megamaser sources \citep{2010ApJ...721...26G, Greene:2016wv} are also presented (with $\ast$ symbol).
Column (4): Apparent luminosity of maser emission (in units of L$_{\odot}$), taken from \citet{2011A&A...532A.125T}, \citet{2009ApJ...695..276B}, \citet{2008ApJ...686L..13G}, and \citet{2008ApJ...685L..39D}. 
Column (5): The Eddington luminosities (Eq. 4) derived from estimating the BH mass, in units of erg\,s$^{-1}$.
Columns (6) \& (7): The Eddington ratio $\dot{m}$ derived from the luminosities at 11\,cm and 6.0\,cm.
Columns (8) \& (9): The mass accretion rates $\dot{M}$ derived from the luminosities at 11\,cm and 6.0\,cm, in units of M$_\odot \rm yr^{-1}$. ($\ast$): For eight sources: UGC3789, NGC1068, NGC1320, NGC2273, NGC3079, NGC3393, NGC4258, NGC4388, the Eddington ratio $\dot{m}$ are also derived from BH mass results taken from \cite{Greene:2016wv}.
\end{flushleft}
\end{table*}

\begin{table*}
\caption[]{The Parameters of Seyfert 2s without H$_2$O maser detections}
\label{para_nonmaser}
\resizebox{\textwidth}{!}{%
\begin{tabular}{lcccccccc}
\hline
Source            & $\sigma$ & $\log$ M$_{\rm BH}$ & UL-H$_2$O & $\log$ \emph L$_{\rm Edd}$ & $\log \dot{m}_{11}$        & $\log \dot{m}_{6.0}$       & $\log \dot{M}_{11}$                      & $\log \dot{M}_{6.0}$                     \\ \hline
NGC 777  & 314 & 8.96 & 1.2  & 47.07 & \textless -8.53 & \textless -8.44 & \textless -7.16  & \textless -7.07  \\
NGC 1058 & 48  & 5.31 & -0.1 & 43.42 & \textless -6.68 & -7.04$\pm$0.1   & \textless -8.96  & -9.32 $\pm$ 0.1  \\
NGC 1358 & 213 & 8.2  & 0.6  & 46.31 & \textless -7.98 & \textless -8.02 & \textless -7.37  & \textless -7.41  \\
NGC 1667 & 170 & 7.76 & 1.1  & 45.87 & -7.07$\pm$0.02  & -7.22$\pm$0.03  & -6.9 $\pm$ 0.02  & -7.05 $\pm$ 0.03 \\
NGC 2655 & 160 & 7.65 & 0.2  & 45.76 & -7.47$\pm$0.03  & -7.5$\pm$0.01   & -7.41 $\pm$ 0.03 & -7.44 $\pm$ 0.01 \\
NGC 2992 & 160 & 7.65 & 0.1  & 45.76 & -6.98$\pm$0.01  & -6.96$\pm$0.01  & -6.92 $\pm$ 0.01 & -6.89 $\pm$ 0.01 \\
NGC 3147 & 225 & 8.31 & 0.7  & 46.42 & -7.96$\pm$0.01  & -7.89$\pm$0.01  & -7.24 $\pm$ 0.01 & -7.17 $\pm$ 0.01 \\
NGC 3185 & 76  & 6.2  & -0.2 & 44.31 & \textless -6.81 & \textless -6.76 & \textless -8.2   & \textless -8.15  \\
NGC 3976 & 191 & 7.99 & 1.1  & 46.10 & -               & -7.86$\pm$0.06  & -         & -7.46 $\pm$ 0.06 \\
NGC 3982 & 70  & 6.04 & -0.4 & 44.15 & -6.6$\pm$0.02   & -6.51$\pm$0.03  & -8.15 $\pm$ 0.02 & -8.05 $\pm$ 0.03 \\
NGC 4138 & 126 & 7.18 & -0.5 & 45.29 & -6.27$\pm$0.02  & -6.65$\pm$0.01  & -6.67 $\pm$ 0.02 & -7.06 $\pm$ 0.01 \\
NGC 4168 & 180 & 7.88 & 0.3  & 45.99 & \textless -8.75 & -8.74$\pm$0.05  & \textless -8.46  & -8.45 $\pm$ 0.05 \\
NGC 4395 & 30  & 4.4  & -1.4 & 42.51 & -6.19$\pm$0.05  & \textless -6.68 & -9.38 $\pm$ 0.05 & \textless -9.87  \\
NGC 4472 & 281 & 8.74 & 0.4  & 46.85 & -8.71$\pm$0.01  & -8.7$\pm$0.01   & -7.56 $\pm$ 0.01 & -7.55 $\pm$ 0.01 \\
NGC 4501 & 167 & 7.73 & -0.4 & 45.84 & -7.68$\pm$0.01  & -7.79$\pm$0.01  & -7.54 $\pm$ 0.01 & -7.64 $\pm$ 0.01 \\
NGC 4565 & 150 & 7.52 & -0.7 & 45.63 & -8.44$\pm$0.01  & -8.58$\pm$0.02  & -8.51 $\pm$ 0.01 & -8.65 $\pm$ 0.02 \\
NGC 4579 & 165 & 7.71 & -0.4 & 45.82 & -7.79$\pm$0.01  & -7.65$\pm$0.01  & -7.67 $\pm$ 0.01 & -7.53 $\pm$ 0.01 \\
NGC 4594 & 231 & 8.36 & 0.4  & 46.47 & -8.37$\pm$0.01  & -8.04$\pm$0.01  & -7.6 $\pm$ 0.01  & -7.27 $\pm$ 0.01 \\
NGC 4725 & 132 & 7.27 & -0.5 & 45.38 & \textless -8.12 & \textless -8.27 & \textless -8.44  & \textless -8.59  \\
NGC 4941 & 98  & 6.7  & -0.4 & 44.81 & \textless -7.24 & \textless -7.52 & \textless -8.13  & \textless -8.41  \\
NGC 5005 & 172 & 7.79 & 0.4  & 45.90 & -7.71$\pm$0.01  & -7.74$\pm$0.01  & -7.51 $\pm$ 0.01 & -7.54 $\pm$ 0.01 \\
NGC 5395 & 146 & 7.47 & 0.7  & 45.58 & -6.99$\pm$0.01  & -7.07$\pm$0.02  & -7.11 $\pm$ 0.01 & -7.19 $\pm$ 0.02 \\
NGC 5899 & - & - & -  & - & -  & -  & - & - \\
NGC 6951 & 115 & 7.01 & -0.1 & 45.12 & -7.13$\pm$0.02  & -7.17$\pm$0.02  & -7.71 $\pm$ 0.02 & -7.74 $\pm$ 0.02 \\
NGC 7743 & 85  & 6.42 & 0.0  & 44.53 & \textless -6.92 & \textless -6.7  & \textless -8.09  & \textless -7.87
              \\ \hline
\end{tabular}
}
\begin{flushleft}
\scriptsize
\textbf{Note:} 
Column (1): Source name. 
Column (2): The same as Col.(2) in Table.\ref{para_maser}. For source NGC 5899, the average stellar velocity dispersion is not available.
Column (3): Blackhole mass, in the units of M$_\odot$, derived from the empirical M-$\sigma$ relation.
Column (4): The same as Col.(5) in Table.\ref{nonmaser}.
Columns (5) to (9): The same as Cols.(5) to (9) in Table.\ref{para_maser}.

\end{flushleft}
\end{table*}




\bibliographystyle{mnras}
\bibliography{Untitled3.bbl} 




\bsp	
\label{lastpage}
\end{document}